\documentclass[%
 reprint,
 amsmath,amssymb,
 aps,
floatfix,
]{revtex4-2}

\usepackage{amsmath}
\usepackage{physics}
\usepackage{graphicx}
\usepackage{dcolumn}
\usepackage{bm}
\usepackage{subfigure}
\usepackage{xcolor}

\begin{document}


\title{Facets of correlated non-Markovian channels}
\author{Vivek Balasaheb Sabale$^{1}$}
\email{sabale.1@iitj.ac.in}
\author{Nihar Ranjan Dash$^{2}$}
\email{dash.1@iitj.ac.in}
\author{Atul Kumar$^{1}$}
\email{atulk@iitj.ac.in}
\author{Subhashish Banerjee$^{2}$}
\email{subhashish@iitj.ac.in}
\affiliation{$^{1}$ Department of Chemistry \\ Indian Institute of Technology Jodhpur, 342030, India}
\affiliation{$^{2}$ Department of Physics \\ Indian Institute of Technology Jodhpur, 342030, India}

\begin{abstract}
We investigate the domain of correlated non-Markovian channels, exploring the potential memory arising from the correlated action of channels and the inherent memory due to non-Markovian dynamics. The impact of channel correlations is studied using different non-Markovianity indicators and measures. In addition, the dynamical aspects of correlated non-Markovian channels, including entanglement dynamics as well as changes in the volume of accessible states, are explored. The analysis is presented for both unital and non-unital correlated channels. A new correlated channel constructed with modified Ornstein-Uhlenbeck noise is also presented and explored. Further, the geometrical effects of the non-Markovianity of the correlated non-Markovian channels are discussed with a study of change in the volume of the accessible states. The link between the correlation factor and error correction success probability is highlighted.
\end{abstract}

\maketitle

\section{\label{sec:level1}Introduction}
The field of quantum information \cite{nielsen2010quantum} is based on utilizing the inherent properties of quantum systems to analyze foundational aspects and to propose efficient applications in diverse academic domains \cite{einstein1935can, bell1964einstein,monz2016realization, boschi1998experimental, peres1997quantum, PhysRevLett.69.2881, haven2002discussion, chen2001quantum, wang2023simulating, singh2022bell, tilly2022variational, tilly2022variational, sharma2023role, sabale}. However, system-environment interaction deteriorates these properties in an open quantum system setting. Interestingly, the system-environment interaction can cause memory effects that lead to a revival of quantum effects. The idea of non-Markovianity \cite{banerjee2018open, rivas2014quantum, breuer2012foundations, PhysRevA.99.042128, milz2021quantum, depolarizing} studies such effects and can lead to a deeper comprehension of the natural world. Memory or non-Markovianity can be used as a resource in quantum information tasks \cite{motavallibashi2021non, PhysRevA.103.052422}. The study of non-Markovianity is further extended to understand complex effects of physical significance in the context of quantum biology \cite{cao2020quantum,lambert2013quantum,huelga2013vibrations}; to enhance the security of continuous-variable quantum key distribution and quantum dialogue \cite{PhysRevA.83.042321,thapliyal2017quantum}; and to study quantum metrology \cite{PhysRevLett.109.233601} and quantum control \cite{PhysRevA.85.032321}. Previous studies of non-Markovian effects were focused on uncorrelated quantum channels \cite{rivas2014quantum,breuer2012foundations,PhysRevA.99.042128,milz2021quantum}. However, quantum channels could be correlated such that there could be correlations between successive channel uses, ending the independence of quantum channels acting on the system state \cite{RevModPhys.86.1203}. These correlated effects of quantum channels are unavoidable for high information transmission rates and have been explored experimentally \cite{PhysRevLett.92.257901}. In this article, we bring forth various tools to study non-Markovian effects in correlated quantum channels to analyze the relation and effect of channel correlation and non-Markovianity. 

The initial exploration of unital non-Markovian correlated quantum channels and their enhanced dynamical non-Markovian memory effects in correlated quantum channels were shown in \cite{PhysRevA.94.032121}, which would not be observed using trace distance-based measures of non-Markovianity. The reported enhancement involving unital quantum channels was observed using an entanglement-based measure in \cite{PhysRevA.94.032121}, which is a state-dependent measure and requires optimization over possible states. Here we undertake the study of unital as well as non-unital correlated non-Markovian channels.  In addition to the entanglement-based measure, here use is also made of interesting identifiers of non-Markovian effects based on the dynamics of the volume of accessible states \cite{gmnonmarkovianity1}, which is state independent. Moreover, using \cite{PhysRevA.65.050301, PhysRevA.94.032121}, we propose and analyze a CP-divisible unital correlated quantum channel with modified Ornstein-Uhlenbeck noise (OUN) \cite{yu2010entanglement,kumar2018non}. In order to quantify the non-Markovianity of correlated OUN channel, we use the Shrikant-Srikanth-Subhashish (SSS) measure based on deviation from temporal self-similarity \cite{utagi2020temporal} which can capture weak non-Markovian as well. The SSS measure helps us comprehend how the channel correlation factor of a correlated non-Markovian CP-divisible channel affects non-Markovianity. Further, we derive a correlated random telegraph noise (RTN) \cite{RTN,kumar2018non} and a correlated non-Markovian amplitude damping (NMAD) channel \cite{nmad1,nmad2}, unital and non-unital in nature, respectively. The associated Kraus operators for fully correlated NMAD are derived using the corresponding master equation \cite{yeo2003time}. For OUN, RTN and NMAD, we further study the signatures of non-Makovianity using state-dependent and state-independent measures. Our results of different types of correlated non-Markovian channels help us to understand the connection between channel correlations and non-Markovianity in a broader sense. Interestingly, we find that increasing the channel correlation further increases the non-Markovianity. In addition to this, we explore the freezing of quantum correlations for a certain class of states. The concurrence of these quantum states shows freezing behaviour in time, which is of interest from a quantum information perspective \cite{Karpat2017,frozen2}. We further studied the performance of the quantum error correction (QEC) \cite{shor1995scheme,calderbank1996good,steane1996error,gottesman1997stabilizer,gottesman2009introduction} under non-Markovian correlated errors. In the context of QEC, use of the concatenated quantum codes \cite{knill1996concatenated,gottesman1997stabilizer,gaitan2008quantum,dash2024concatenating} is the first step to realize fault-tolerant quantum computation \cite{shor1997fault, gottesman2009introduction}. We use the concatenated quantum error correcting codes for correlated unital channels under non-Markovian dynamics. In particular, we study error correction for correlated RTN and correlated OUN noise channels. \par

The work is organized as follows: In Sec. \ref{sec:level2}, a general idea of quantum channels is given, with additional discussion of correlated action channels. The constructed channels and their dynamic maps are presented in Sec. \ref{sec:level3}. In Sec. \ref{sec:level4}, we present a study of non-Markovianity identifiers based on concurrence and the SSS measure. The state-independent identifier of non-Markovinity based on the volume of physical states accessible to the system affected by correlated quantum channel is presented in Sec. \ref{sec:level4iv}. In Sec. \ref{sec:level5i}, the discussion about the freezing of correlations with the increase in channel correlation factor is presented. Further, the concatenated code is used to study the quantum error correction under correlated unital channel discussed in Sec. \ref{sec:level5}. The Sec. \ref{sec:level6} concludes the article with results and summary.

\section{\label{sec:level2} Quantum channels and their correlated action}
Quantum channels \cite{RevModPhys.86.1203} map the evolution of a quantum state from initial state $\rho$ to the final state $\rho \rightarrow \mathcal{E}(\rho)$. For the case of a single qubit density operator $\rho$, the dynamical evolution is given by 
\begin{equation}
    \rho \rightarrow \mathcal{E}(\rho) = \sum_{i=0}^{n}A_{i} \rho A_{i}^{\dagger},
    \label{eq: channel}
\end{equation}
where $A_{i}$ are single qubit Kraus operators \cite{kraus1983states} corresponding to a quantum channel under study. \par
Interestingly, for a two-qubit system, quantum channels can be classified as uncorrelated as well as correlated channels \cite{caruso2014quantum}, where the uncorrelated channel is an extension of a single qubit channel through the tensor product $\mathcal{E}^{\otimes2}=\mathcal{E}\otimes \mathcal{E}$. The following equation represents an uncorrelated quantum channel map 
\begin{equation}
    \rho \rightarrow \mathcal{E}^{\otimes2}(\rho) = \sum_{i,j=0}^{n}A_{i}A_{j} \rho A_{i}^{\dagger}A_{j}^{\dagger}.
    \label{eq: uncorrelated channel}  
\end{equation}
Channels that defy the logic of tensorial product are called correlated quantum channels \cite{yeo2003time}. The correlation in such channels can arise from correlated noise probabilities or correlated Kraus operators. 
The action of a correlated quantum channel can be described as 
\begin{equation}
    \rho \rightarrow \mathcal{E}^{corr}(\rho) = \sum_{i_{1}i_{2}}A_{i_{1}i_{2}} \rho A_{i_{1}i_{2}}^{\dagger},
    \label{eq: correlated channel} 
\end{equation}
where $A_{i_{1}i_{2}}=\sqrt{p_{i_{1}i_{2}}} B_{i_{1}}\otimes B_{i_{2}}$ are Kraus operators acting on the two-qubit density operator $\rho$, the sum of joint probabilities is 1 ($\sum_{i_{1}i_{2}} p_{i_{1}i_{2}} = 1$). The expression for joint probabilities is given as
\begin{equation}
        p_{i_{1}i_{2}}= (1- \mu) p_{i_{1}}p_{i_{2}} + \mu p_{i_{i_{1}}} \delta_{i_{1}i_{2}},
         \label{eq:jointprob1}
\end{equation}
where $\mu$ is the channel correlation factor signifying the degree of classical correlation in channel actions, which ranges from $0$ to $1$. For $\mu=0$, joint probabilities are separable, indicating the independence of noise actions on two qubits and their uncorrelated nature. A channel is fully correlated for the case $\mu=1$, implying the complete dependence of the channels action on both the qubits. For a nonzero value of $\mu$, the channel is correlated satisfying $\mathcal{E}^{corr} \neq \mathcal{E}\otimes \mathcal{E} $.
The Kraus operator form of correlated quantum channels is
\begin{equation}
\begin{split}
\mathcal{E}^{corr} (\rho) & = (1-\mu)\sum_{i_{1}i_{2}}A_{i_{1}i_{2}} \rho A_{i_{1}i_{2}}^{\dagger} + \mu \sum_{j}A_{jj} \rho A_{jj}^{\dagger} \\
 & =(1-\mu)\mathcal{E}^{uncorr} (\rho) + \mu \mathcal{E}^{fcorr}(\rho),
\end{split}
\label{corr_gen_map}
\end{equation}
where $\mathcal{E}^{fcorr}$ is map of fully correlated quantum channel and $\mathcal{E}^{uncorr}$ is uncorrelated quantum channel for which $P_{i_{1}i_{2}}= p_{i_{1}}p_{i_{2}}$. \par
In this article, we analyze correlated channels belonging to unital ($\mathcal{E}_{u}$) and non-unital ($\mathcal{E}_{nu}$) classes. The unital quantum channels are known to preserve the identity operator and satisfy the relation $\mathcal{E}_{u}(\mathbf{I}_{2})=\mathbf{I}_{2}$, the non-unital channels do not satisfy this condition, i.e. ($\mathcal{E}_{nu}(\mathbf{I}_{2})\neq \mathbf{I}_{2}$) and hence do not preserve the identity operator. \par

\section{\label{sec:level3} Correlated non-Markovian channels}
The correlated non-Markovian channels can be classified into two classes, namely unital and non-unital. The conditions for the classification of these channels are described in the previous section. The model proposed in \cite{PhysRevA.65.050301} was used to construct correlated unital channels. Although there are a few instances of characterizing unital correlated non-Markovian channels \cite{PhysRevA.94.032121,Guo_2023}, studying other types of correlated non-Markovian channels is necessary to comprehend the relationship between non-Markovianity and channel correlation factor $\mu$. In addition to the study in \cite{PhysRevA.94.032121}, which looked at correlated RTN of a CP-indivisible nature, we also study correlated non-Markovian channels that are CP-divisible. It is also imperative to study and analyze non-unital correlated non-Markovian channels, as such channels further lead to errors related to energy dissipation in addition to dephasing. In the next subsection, we demonstrate the constructions of unital as well as correlated non-unital non-Markovian channels.

\subsection{\label{sec:level3i} Correlated unital non-Markovian channel}
This subsection discusses unital CP-indivisible correlated RTN and unital CP-divisible correlated OUN channels. The single-qubit Kraus operators  used to construct two-qubit correlated unital non-Markovian channels are expressed as
\begin{equation}
    K_{i}= \sqrt{q_{i}} \sigma_{i},
    \label{eq:RTN_op}
\end{equation}
where $i=0, 1, 2, 3$ and $\sigma_{i}$ are Pauli operators. Here, the coefficients $q_{i}$ associated with Kraus operators of RTN and OUN channels \cite{yu2010entanglement,kumar2018non} are of the form 
\begin{equation}
    q_{0}= \frac{1}{2}[1+p(t)] , \quad q_{1}=q_{2}=0,\quad q_{3}= \frac{1}{2}[1-p(t)].
    \label{eq:prob}
\end{equation}
The function $p(t)$ in Eq. (\ref{eq:prob})  represents a noise function given by
\begin{equation}
    p(t)= 
\begin{cases}
    \exp(-\gamma t)(\cos{(\omega \gamma t)}+\frac{\sin{(\omega \gamma t)}}{\omega}) & \text{for RTN}, \\
    \\
    \exp[-\frac{G}{2}(t+\frac{1}{g}(\exp(-g t)-1))]              & \text{for OUN}.
\end{cases}
 \label{eq:probtime}
\end{equation}
Here, $\omega=\sqrt{(2a/\gamma)^{2}-1}$ denotes the noise parameter of RTN where $a$ signifies  strength of system-environment coupling and $\gamma$ corresponds to the fluctuation rate of RTN. Similarly, for OUN, $g^{-1}=\tau_{c}$ is the finite correlation time of the environment. The parameter $G$ is the effective relaxation time, frequently called $T_{1}$ in the literature \cite{kumar2018non}.  \par
Using Eq. (\ref{eq: uncorrelated channel}), the map corresponding to uncorrelated quantum channel leads to 
\begin{eqnarray}
    \mathcal{E}^{uncorr}(\rho) = \sum_{ij}q_{i}q_{j}(\sigma_{i}\otimes \sigma_{j})\rho (\sigma_{i}\otimes \sigma_{j}),
    \label{eq: uncorr_map}
\end{eqnarray}
implying no correlation between successive uses of channels.
Our analysis for the construction of correlated channels will now lead to correlated errors on qubits with joint probability. Using Eq. (\ref{eq:jointprob1}), the joint probability is given as \begin{equation}
        p_{ij}= (1- \mu) q_{i}q_{j} + \mu q_{i} \delta_{ij},
         \label{eq:jointprob}
\end{equation}
where $\mu$ is the channel correlation factor that controls the joint probability and, hence, the channel correlations and induced errors. The factor $\mu$ accounts for memory effects due to classical correlation between successive uses of the quantum channel \cite{sk2022protecting}. The maps corresponding to correlated RTN and OUN channels can be represented as
\begin{multline}
       \mathcal{E}^{corr}(\rho)= p_{00}(\sigma_{0}\otimes \sigma_{0})\rho(\sigma_{0}\otimes \sigma_{0})\\ 
    + p_{03}(\sigma_{0}\otimes \sigma_{3})\rho(\sigma_{0}\otimes \sigma_{3}) \\
    + p_{30}(\sigma_{3}\otimes \sigma_{0})\rho(\sigma_{3}\otimes \sigma_{0})\\
    + p_{33}(\sigma_{3}\otimes \sigma_{3})\rho(\sigma_{3}\otimes \sigma_{3}). \\
    \label{eq:corr_map_oun_rtn}
\end{multline}
The form of $p_{ij}$ can easily be obtained from Eq. (\ref{eq:jointprob}), using Eqs. (\ref{eq:prob}) and (\ref{eq:probtime}). As $p_{ij}$ depends on $\mu$, the increase in $\mu$ leads to correlated errors on two qubits, and for $\mu=1$, the channel becomes fully correlated. 

\begin{figure*}
    \subfigure[]{\includegraphics[width=0.33\linewidth]{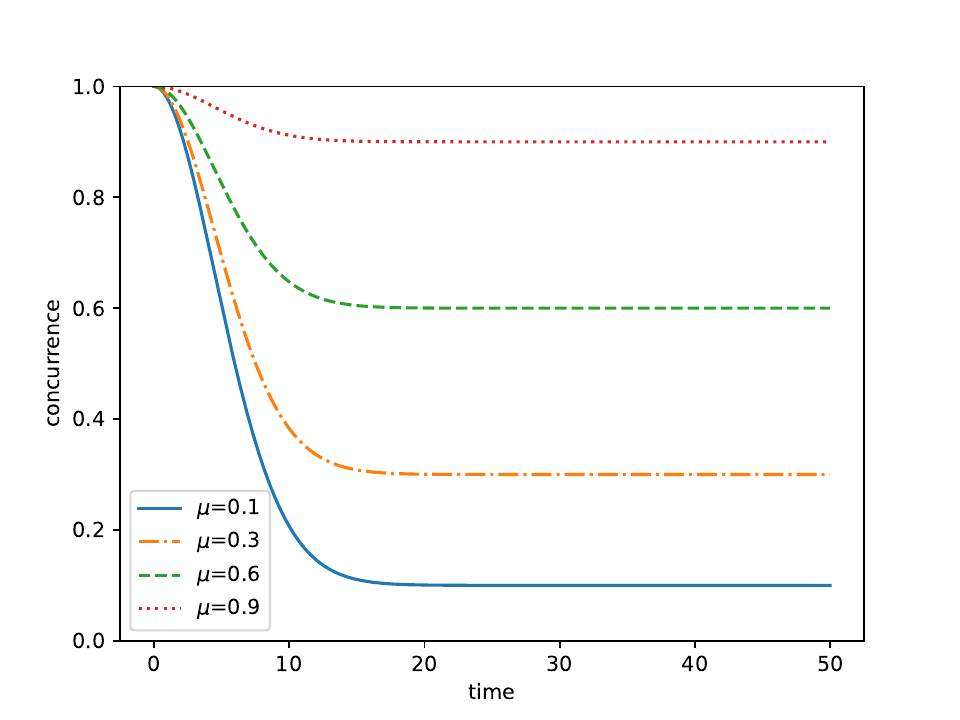}}
    \subfigure[]{\includegraphics[width=0.33\linewidth]{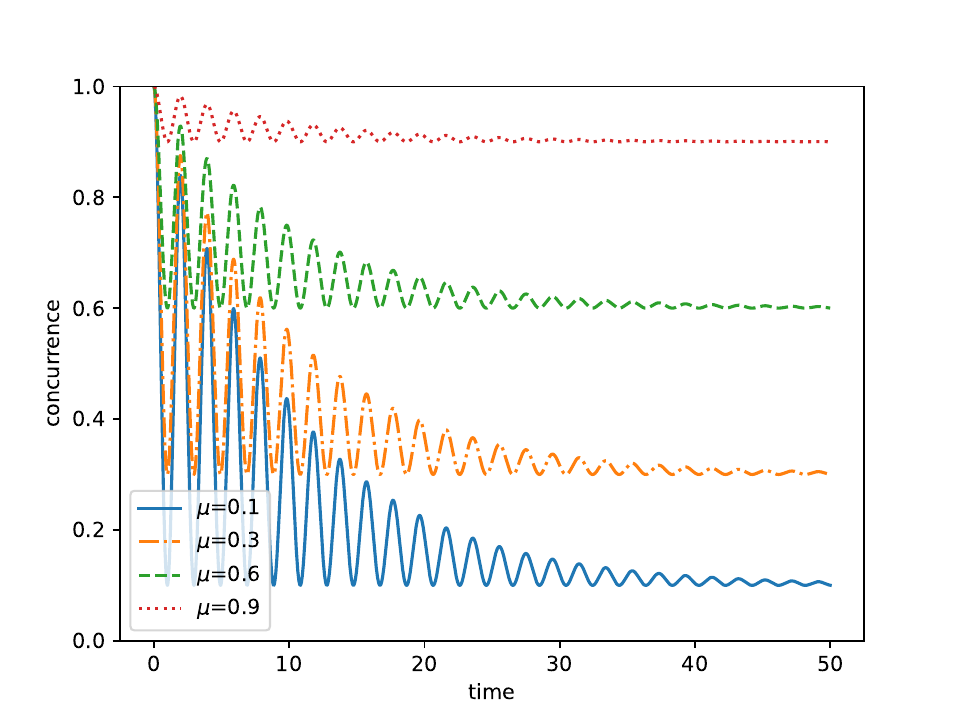}}\hfill
    \subfigure[]{\includegraphics[width=0.33\linewidth]{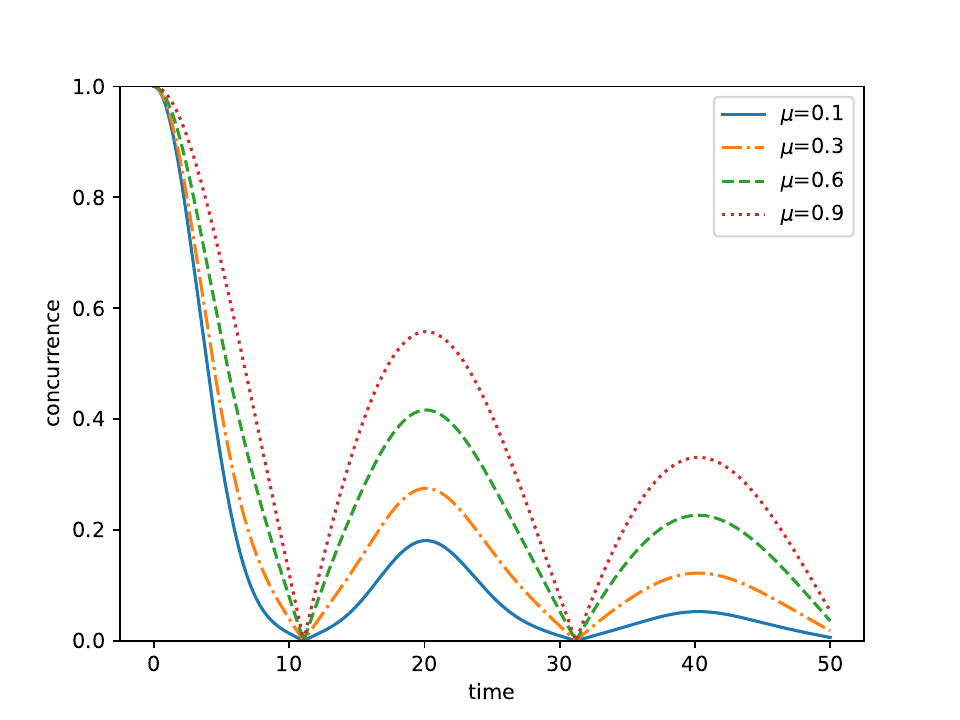}}\hfill
    \subfigure[]{\includegraphics[width=0.33\linewidth]{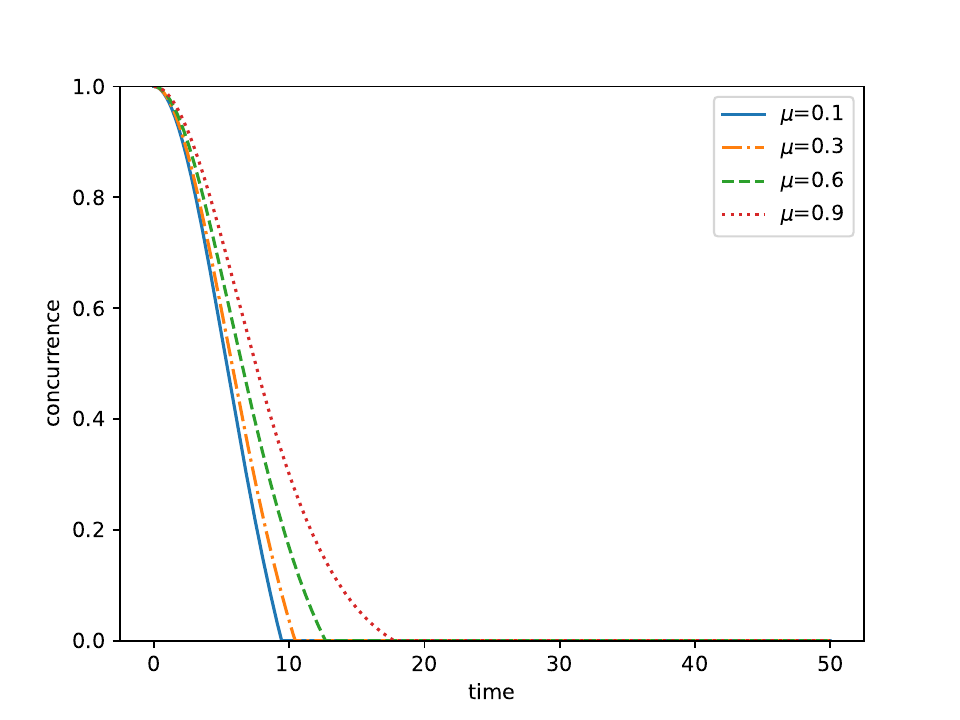}}\hfill 
    \subfigure[]{\includegraphics[width=0.33\linewidth]{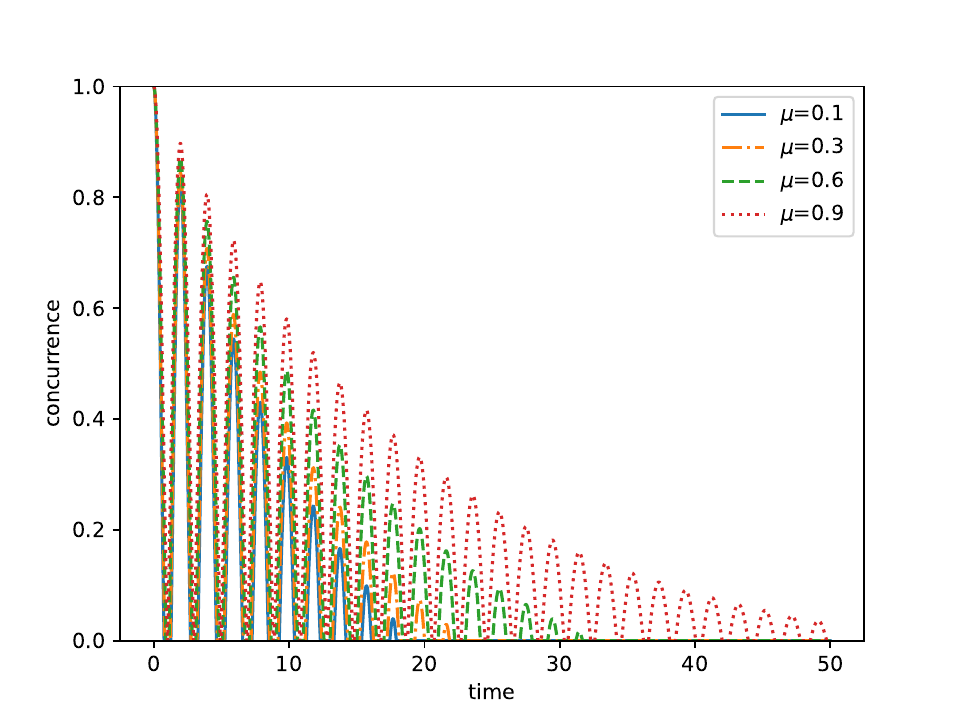}}\hfill
    \subfigure[]{\includegraphics[width=0.33\linewidth]{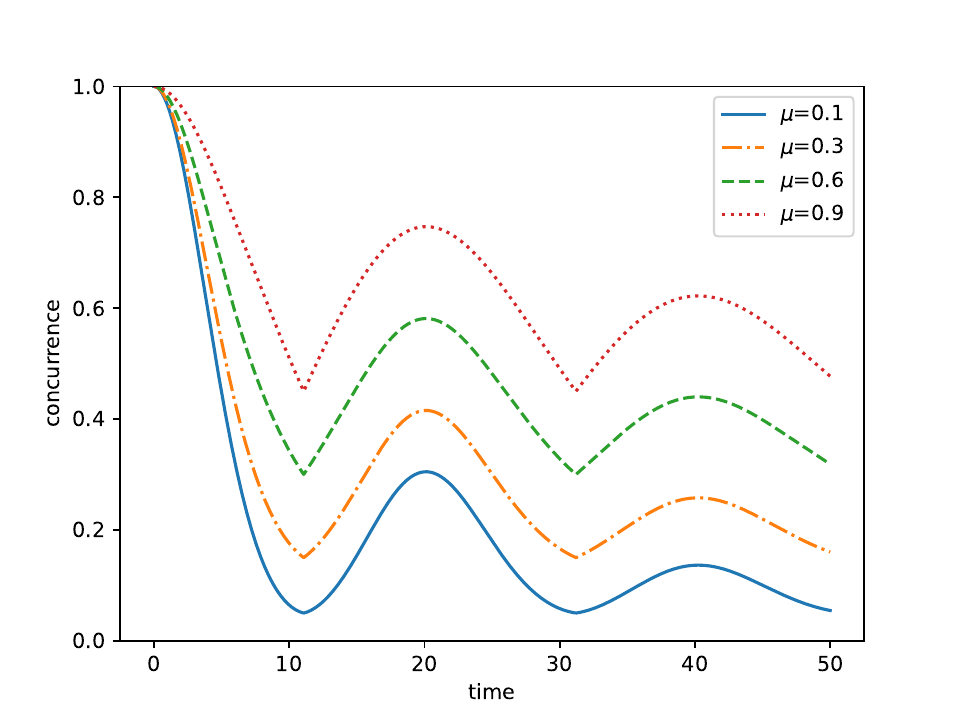}}\\
    \caption{(a) Evolution of state $\ket{\phi^{\pm}}=\frac{1}{\sqrt{2}} [\ket{00} \pm \ket{11} ]$ in correlated-OUN channel ($G=1, g=0.05$) (b) Evolution of state $\ket{\phi^{\pm}}=\frac{1}{\sqrt{2}} [\ket{00} \pm \ket{11} ]$ in correlated RTN channel ($a=0.8, \gamma=0.05 $) (c) Evolution of state $\ket{\phi^{\pm}}=\frac{1}{\sqrt{2}} [\ket{00} \pm \ket{11} ]$ in correlated NMAD channel ($g=0.05, \gamma_{0}=1$) (d)Evolution of state $\ket{\alpha}=\frac{1}{2}[\ket{00}+\ket{01}+\ket{10}-\ket{11}]$ in correlated OUN channel ($G=1, g=0.05$)  (e) Evolution of state $\ket{\alpha}=\frac{1}{2}[\ket{00}+\ket{01}+\ket{10}-\ket{11}]$ in correlated RTN channel ($a=0.8, \gamma=0.05 $) (f) Evolution of state $\ket{\alpha}=\frac{1}{2}[\ket{00}+\ket{01}+\ket{10}-\ket{11}]$ in correlated NMAD channel ($g=0.05, \gamma_{0}=1$).}
 \label{fig: OUN_RTN_NMAD}
\end{figure*}

\subsection{\label{sec:level3ii} Correlated non-unital non-Markovian channels}
The non-Markovian amplitude damping channel \cite{nmad1,nmad2} belongs to the non-unital class. The master equation for a non-Markovian amplitude damping channel is 
\begin{equation}
    \mathcal{L}(\rho)= \gamma(t)[\sigma_{-}\rho \sigma_{+} 
    -\frac{1}{2} \sigma_{+}\sigma_{-}\rho -\frac{1}{2} \rho \sigma_{+}\sigma_{-}],
    \label{eq:one_qubit_nmad_me}
\end{equation}
where $\sigma_{\pm}$ are raising and lowering operators, which are defined as $\sigma_{\pm}= \frac{1}{2}(\sigma_{x} \pm i \sigma_{y})$. The term $\gamma(t)$ represents a time-dependent decay rate and is given by 
\begin{equation}
    \gamma(t)= -\frac{2}{\abs{G(t)}}\frac{ d\abs{G(t)}}{dt},
\end{equation}
where, $G(t)$ is decoherence function expressed as 
\begin{equation}
    G(t)= e^{-\frac{gt}{2}}\left (\cosh{\left[\frac{lt}{2}\right]} + \frac{g}{l}\sinh{\left[\frac{lt}{2}\right]}\right),
\end{equation}
and $l=\sqrt{g^{2}-2\gamma_{0}g}$. The Eq. (\ref{eq:one_qubit_nmad_me}) can be further extended to a two-qubit system \cite{corr_ad}. It will involve the correlated action of the channel leading to non-unital correlated NMAD. The master equation for a two-qubit NMAD channel will take the following form
\begin{multline}
    \mathcal{L}^{fcorr}(\rho) = \gamma(t) [( \sigma_{-} \otimes \sigma_{-}) \rho (\sigma_{+} \otimes \sigma_{+})\\
    - \frac{1}{2}(\sigma_{+} \otimes \sigma_{+})( \sigma_{-} \otimes \sigma_{-}) \rho \\
    -\frac{1}{2} \rho (\sigma_{+} \otimes \sigma_{+}) ( \sigma_{-} \otimes \sigma_{-})].\\
    \label{eq:two_qubit_nmad_me}
\end{multline}
One can evaluate the map $\mathcal{E}^{fcorr}$ corresponding to the above master equation and corresponding Kraus operators. Following the procedure in Appendix A, we derive Kraus operators for a fully correlated NMAD channel corresponding to the map $\mathcal{E}^{fcorr}$ such that
\begin{subequations}
\label{eq:whole}
\begin{eqnarray}
E_{00}=
\begin{pmatrix}
1 & 0 & 0 & 0\\
0 & 1 & 0 & 0\\
0 & 0 & 1 & 0\\
0 & 0 & 0 & \sqrt{1-p(t)}
\end{pmatrix} ,\label{subeq:E0}
\\\
E_{11}=
\begin{pmatrix}
0 & 0 & 0 & \sqrt{p(t)}\\
0 & 0 & 0 & 0\\
0 & 0 & 0 & 0\\
0 & 0 & 0 & 0
\end{pmatrix},\label{subeq:E1}
\end{eqnarray}
\label{eq:CNMAD_Kraus}
\end{subequations}
where $p(t)$ is the noise function of the NMAD channel and is given by
\begin{equation}
    p(t)=1-\abs{G(t)}^{2}.
    \label{eq:pt_nmad}
\end{equation}
\par
The correlated amplitude damping channel is then expressed using Eq. (\ref{corr_gen_map}), where the superscript $uncorr$ represents a two-qubit uncorrelated NMAD noise and $fcorr$ represents a two-qubit fully-correlated NMAD noise. The uncorrelated part of the channel ($\mathcal{E}^{uncorr}$) mentioned above is expressed using the following equation
\begin{equation}
     \mathcal{E}^{uncorr}(\rho)= \sum_{i,j}(A_{i} \otimes A_{j}) \rho (A_{i}^{\dagger} \otimes A_{j}^{\dagger}).
\end{equation}
Where $A_{i}$ are Kraus operators of NMAD channel such that
\begin{subequations}
\label{eq:whole1}
\begin{eqnarray}
    A_{0}=\begin{pmatrix}
    1 & 0\\
    0 & \sqrt{1-p(t)} 
    \end{pmatrix} ,\label{subeq:a3}
\\\
    A_{1}=
    \begin{pmatrix}
    0 & \sqrt{p(t)}\\
    0 & 0 
    \end{pmatrix}.\label{subeq:a4}
\end{eqnarray}
 \label{eq:NMAD_sq}
\end{subequations}
The fully correlated ($\mathcal{E}^{fcorr}$) part in Eq. (\ref{corr_gen_map}) is given by 
\begin{equation}
    \mathcal{E}^{fcorr}(\rho)= \sum_{i}E_{ii} \rho E_{ii}^{\dagger},
\end{equation}
Here $E_{ii}$ are two-qubit Kraus operators in Eqs. (\ref{subeq:E0}) and (\ref{subeq:E1}).

\section{\label{sec:level4} Identification and Quantification of effects of $\mu$ on non-Markovianity}

The correlated errors played a significant role in the classical Shannon information theory \cite{PhysRevA.65.050301}. The assumption of considering an information transmission channel as correlated leads to beneficial results in tackling noise. The use of such correlated channels allowed one to overcome the noisy effects. Inspired by the idea of non-Markovianity as a resource and the applicability of correlated channels in error suppression, we explore the domain of correlated non-Markovian channels. We also aspire to connect the non-Markovianity and correlation factor $\mu$ of channels. To investigate non-Markovianity, we study the variation of concurrence of a maximally entangled state with time. Further, to quantify non-Markovianity of correlated OUN, we use the SSS measure \cite{utagi2020temporal}. 

\subsection{\label{sec:level4i} Using concurrence variation and revival rates}
The non-monotonic behaviour in a variation of trace distance between two states undergoing dynamics is the easiest way to identify the non-Markovinity or information backflow \cite{laine2010measure}. For states $\rho_{1}(t)$ and $\rho_{2}(t)$, the trace distance is defined by 
\begin{equation}
D(\rho_{1}(t),\rho_{2}(t))=\frac{1}{2} tr\abs{\rho_{1}(t)-\rho_{2}(t)},
\end{equation}
where $\abs{A}=\sqrt{A^{\dagger}A}$. For non-Markovian dynamics, we see a violation of the condition 
\begin{equation}
    D(\rho_{1}(t),\rho_{2}(t)) \leq  D(\rho_{1}(s),\rho_{2}(s)), 0\leq s \leq t,
\end{equation}
due to information backflow happening at some later time.
This implies that if $d D(t)/ d t > 0$ the dynamics is non-Markovian. The non-Markovianity, in this scenario, can be quantified by using the BLP measure \cite{laine2010measure} as given by 
\begin{equation}
    \mathcal{N_{D}(E)} = \underset{\rho_{1}(0),\rho_{2}(0)}{max} \int_{d D(t)/ d t > 0} \frac{d D(t)}{d t}dt.
\end{equation}

\begin{figure}
    \subfigure[]{\includegraphics[width=0.95\linewidth]{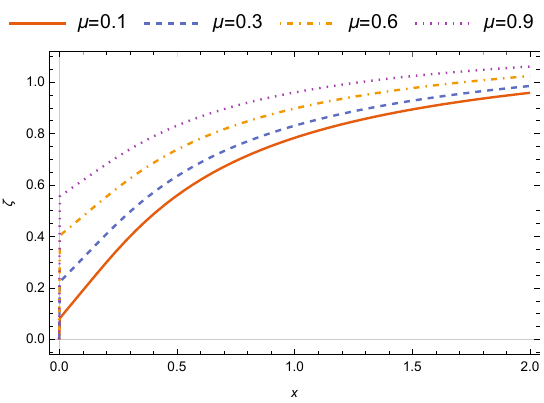}}
    \caption{SSS Measure for correlated non-Markovian OUN channel ($G=0.6$). The x in the figure represents $g^{-1}$. }
 \label{fig: SSS_OUN}
\end{figure}

\begin{figure*}[]
    \subfigure[]{\includegraphics[width=0.3\linewidth]{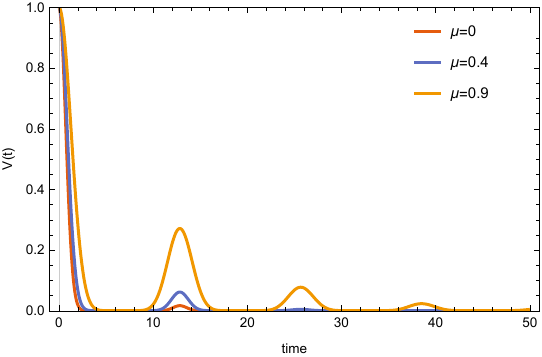}}\hfill
    \subfigure[]{\includegraphics[width=0.3\linewidth]{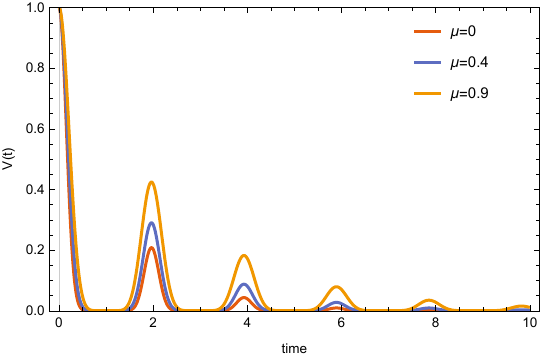}}\hfill
    \subfigure[]{\includegraphics[width=0.3\linewidth]{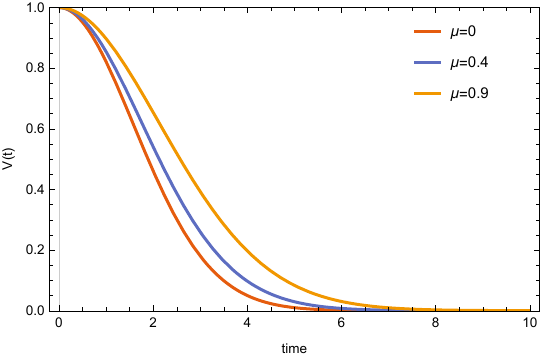}}
    \caption{The variation in volume of physical states ($V(t)=det[F(t)])$, (a) for correlated-NMAD channel ($g=0.02, \gamma_{0}=6$) (b) correlated-RTN channel ($a=0.8, \gamma=0.05 $) (c) correlated-OUN channel ($g=0.05, G=1$). }
 \label{fig: Accessible_volume}
\end{figure*}

One can also detect non-Markovianity by exploiting properties like entanglement, mutual information, or some other quantum information properties \cite{PhysRevLett.105.050403,MI,coherence}. For that, we introduce an ancillary system with the same dimension as the system under investigation. We assume the dynamical map $\mathcal{E}_{t}$ is acting on the system, and the ancillary system evolves trivially. In the absence of any memory or non-Markovian effects, one of the systems evolves without affecting the ancillary system. This points to a condition that is violated by non-Markovian dynamics
\begin{equation}
    X[(I\otimes \mathcal{E}_{t}) \rho_{SA}] \leq X[(I\otimes \mathcal{E}_{s}) \rho_{SA}], 0\leq s \leq t.
    \label{eq:neq_x}
\end{equation}
Here, $\rho_{SA}$ is a density operator associated with the combined system-ancilla setting, and X is any quantum information property that follows monotonic behaviour for a completely positive and trace preserving (CPTP) map. The violation of the above inequality is interpreted as a deviation from divisibility of the dynamical map ($\mathcal{E}_{t}=\mathcal{E}_{t,s}\mathcal{E}_{s}$) and points that the intermediate map $\mathcal{E}_{t,s}$ is not CPTP. Due to the addition of ancillary qubits, this approach differs from the trace distance-based approach. 
\par
For a correlated noise case, We will need a system of at least two qubits. Further, one needs two ancillary qubits to verify the above inequality in case of two qubit-correlated noise channels, making the state under investigation a four-qubit state. This will be hard to investigate with increased Hilbert space dimensions and the involved optimization needed for most measures of non-Markovianity.
The simplification of Eq. (\ref{eq:neq_x}) for a two-qubit system without any ancilla results in the following equation
\begin{equation}
        X[ \mathcal{E}_{t} \rho_{AB}] \leq X[\mathcal{E}_{s} \rho_{AB}],
    \label{eq:neq_x_c}
\end{equation}
for $0\leq s \leq t$ and $\rho_{AB}$ corresponds to the system density matrix \cite{PhysRevA.94.032121}. Following \cite{PhysRevA.94.032121}, we can quantify the memory effects using the following equation
\begin{equation}
    \mathcal{N(E)} = \underset{\rho_{AB}}{max} \int_{d X(t)/ d t > 0} \frac{d X(t)}{d t}dt.
    \label{eq:con_measure}
\end{equation}
As the correlated map $\mathcal{E}_{t}$ is no longer local due to correlated errors on qubits involved, we cannot use the monotonous nature of certain quantum properties like mutual information to detect the violation of divisibility. The correlated maps can be implemented using local operations and classical communication (LOCC). We then resort to entanglement quantifier $X$ to detect a violation of divisibility, as the entanglement is monotonic under LOCC. This further points out that the condition in Eq. (\ref{eq:neq_x_c}) is violated only for non-divisible dynamics. \par
For this, we study the variation of concurrence $C(t)$, an entanglement quantifier $X$, for both unital and non-unital non-Markovian channels for different maximally entangled states.
The concurrence is defined as
\begin{equation}
    C(\rho)=\max\{ 0, \lambda_{1}-\lambda_{2}-\lambda_{3}-\lambda_{4}\},
\end{equation}
where $\lambda_{1},\lambda_{2},\lambda_{3},\lambda_{4}$ are square roots of eigenvalues of the matrix $\rho \Tilde{\rho}$; $\Tilde{\rho}=(\sigma_{y}\otimes\sigma_{y})\rho^{*}(\sigma_{y}\otimes\sigma_{y})$ and $\rho^{*}$ is complex conjugate of a $\rho$. The states used in the study are generated through local unitary operations on Bell states. The revival rates of concurrence are state-dependent. In Fig. \ref{fig: OUN_RTN_NMAD}, we see a decrease in concurrence revival rates ($dC(t)/d t$) for the Bell state evolving under correlated unital channel. In contrast, we observe an increment in concurrence revival rates for the correlated non-unital channels. This suggests that the study of non-Markovian effects, based on measure Eq. \ref{eq:con_measure}, would need optimization and hence different states for non-unital or unital correlated channels. This is depicted in Figs. \ref{fig: OUN_RTN_NMAD} (b), (c), (e), (f). Under the influence of unital correlated channels, the dynamics of the Bell state starts to freeze for a higher correlation factor $\mu$. This can be seen from Fig. \ref{fig: OUN_RTN_NMAD} (a) and (b). The state resulting from the local unitary transformation of the Bell state exhibits enhanced non-Markovianity, as depicted in Fig. \ref{fig: OUN_RTN_NMAD} (d) and (e). There are two types of non-Markovianity, one by retaining correlation, as in the case of correlated OUN \cite{kumar2018non} and depicted in Figs. \ref{fig: OUN_RTN_NMAD} (a) and (d), and the other by information backflow observed for correlated RTN and NMAD as shown in Figs. \ref{fig: OUN_RTN_NMAD} (b), (c), (e) and (f); in both cases, we see enhanced non-markovianity with increase in channel correlation factor $\mu$. In the case of the non-unital correlated NMAD channel, we observed an enhanced revival rate for the Bell state, indicating enhanced non-Markovianity. It is depicted in Fig. \ref{fig: OUN_RTN_NMAD} (c). This further motivates us to identify and quantify increased non-Markovianity due to channel correlation with new concepts.

\subsection{\label{sec:level4iii} Using SSS Measure based on temporal self-similarity for correlated OUN channel}
The study of the dynamics of concurrence points to a prolongation of concurrence decay time with increased $\mu$ for correlated OUN channels. The absence of a revival rate of concurrence in the case of correlated OUN channel points towards the need for another measure that captures this type of non-Markovianity arising from CP-divisible dynamics. The idea of deviation from temporal self-similarity can handle the above-presented issues. One can use the SSS measure based on temporal self-similarity to study non-Markovian CP-divisible processes. \par
Temporal self-similarity, the central idea behind the SSS measure, can be understood in terms of the independence of the intermediate dynamical map from the initial time $t_{0}$. The dynamics arising in such a case is free from the history of effects arising from the system-environment interaction during evolution. The requirement for information about the initial time $t_{0}$ points to the memory effects in dynamics; these effects are quantified with the help of SSS measure of non-Markovianity.\par
The  SSS measure can be computed as
\begin{equation}
    \zeta = min_{\mathcal{L}^{*}} \frac{1}{T}\int_{0}^{T}\norm{\mathcal{L}(t)-\mathcal{L}^{*}} dt,
    \label{eq:SSS_measure}
\end{equation}
where $\mathcal{L}^{*}$ is a time-independent generator of the corresponding quantum dynamical semigroup. We use the procedure from Appendix A to compute $\mathcal{L}(t)$ and corresponding $\mathcal{L}^{*}$ for correlated non-Markovian OUN quantum channels with the corresponding Markov limit obtained as $\frac{1}{g}\rightarrow 0$.\par
The results for the correlated OUN channel show that an increase in correlation factor $\mu$ leads to an increase in non-Markovinity $\zeta$. The results are depicted in Fig. \ref{fig: SSS_OUN}, which shows the dependence of non-Markovianity on the correlation factor $\mu$. In addition, we show the effect of channel correlation factor on quantum error correction. This will be discussed in Section \ref{sec:level5}. \par

\begin{figure*}
    \subfigure[]{\includegraphics[width=0.33\linewidth]{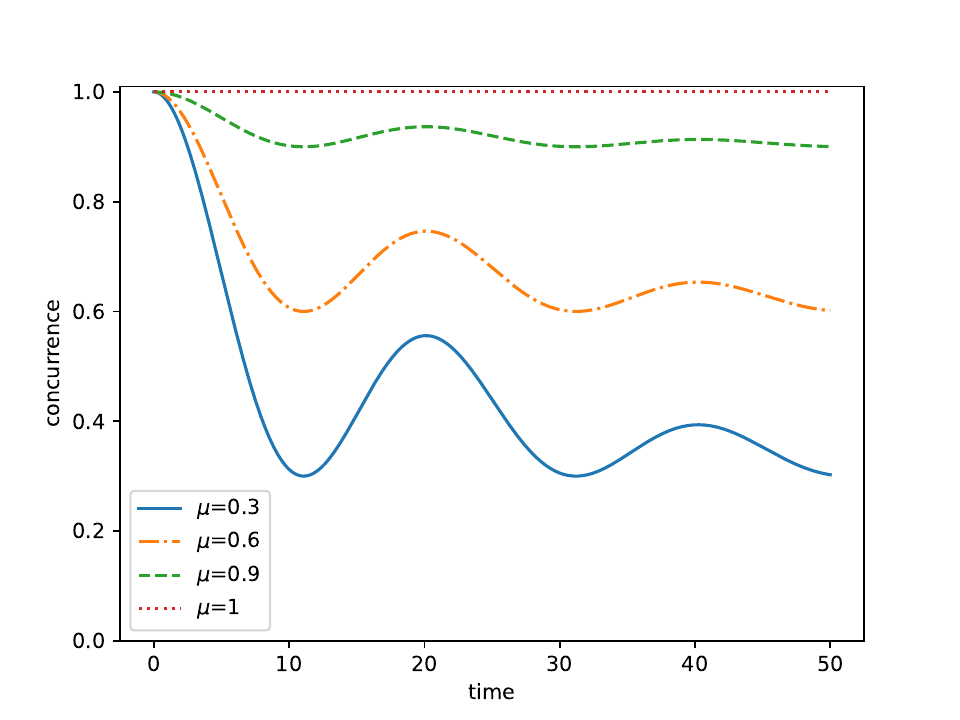}}
    \subfigure[]{\includegraphics[width=0.33\linewidth]{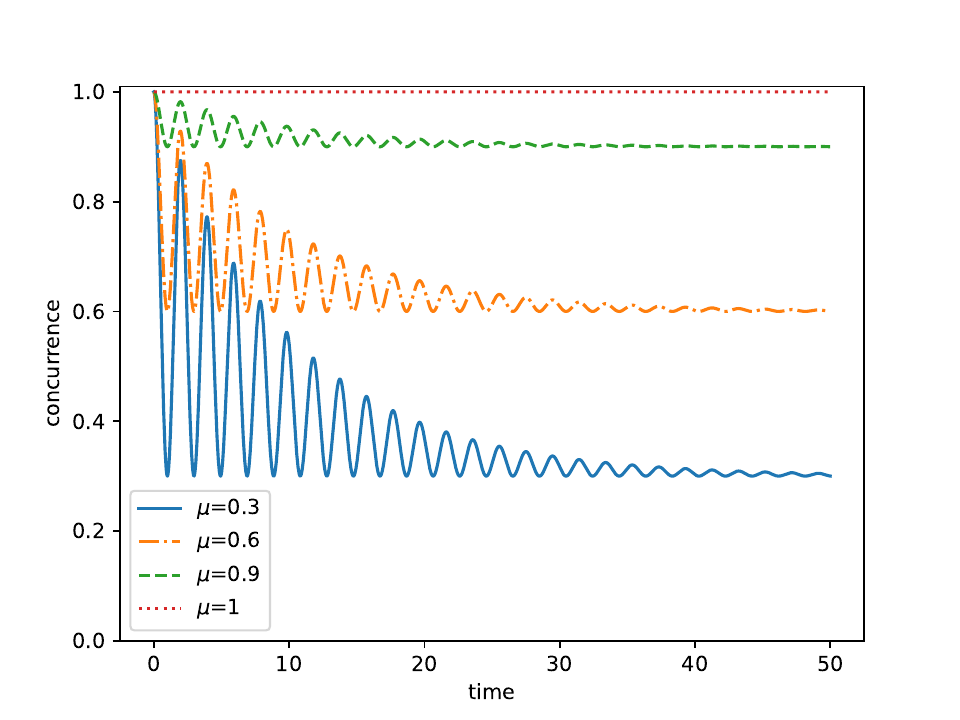}}\hfill
    \subfigure[]{\includegraphics[width=0.33\linewidth]{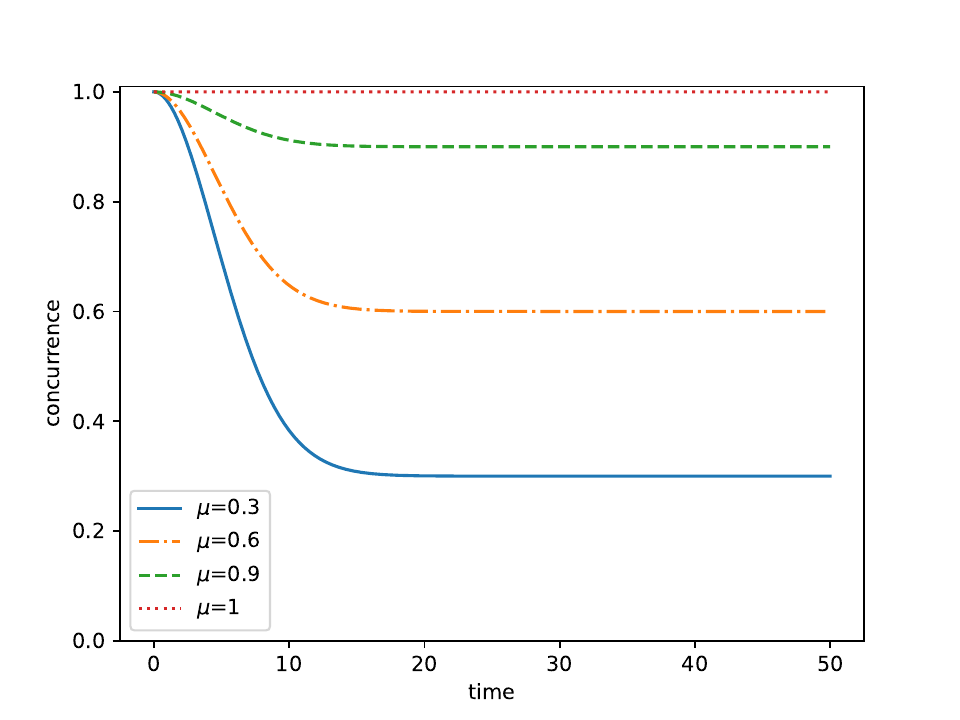}}\hfill
    \caption{(a) Evolution of state $\ket{\psi^{\pm}}=\frac{1}{\sqrt{2}} [\ket{01} \pm \ket{10} ]$ in correlated-NMAD channel ($g=0.05, \gamma_{0}=1$) (b) Evolution of state $\ket{\phi^{\pm}}=\frac{1}{\sqrt{2}} [\ket{00} \pm \ket{11} ]$ in correlated-RTN channel ($a=0.8, \gamma=0.05 $) (c) Evolution of state $\ket{\phi^{\pm}}=\frac{1}{\sqrt{2}} [\ket{00} \pm \ket{11} ]$ in correlated-OUN channel ($g=0.05, G=1$). }
 \label{fig: Freezing}
\end{figure*}

\section{\label{sec:level4iv} Volume of accessible states and its dependence on $\mu$}

In this section, we demonstrate the variation of geometrical volume of the physical states of the system affected by correlated non-Markovian quantum channels. The effect of the channel correlation factor ($\mu$) on the volume of physical states is studied. Furthermore, an analysis of non-Markovian effects is presented, bringing forth a geometrical interpretation and state-independent measure of non-Markovianity studied previously in \cite{PhysRevA.94.032121}. \par

The idea about the dynamics of the volume of physical states accessible to the system can give interesting insights into correlated non-Markovian channel behaviour. For this, one can use a dynamical map, represented by a matrix $\textbf{F(t)}$, associated with a quantum channel. We resort to the Appendix and find the $\textbf{F(t)}$ matrix corresponding to the correlated channels under study (see Eq. \ref{eq:F_mat}). The determinant of the matrix $\textbf{F(t)}$ provides insight into the system dynamics; it gives an idea about the volume of states accessible to the system. 
The dynamical variation of the volume of accessible states is given by \cite{gmnonmarkovianity1,gmnonmarkovianity2}
\begin{equation}
    V(t)= det[F(t)].
    \label{eq:volume_accesible}
\end{equation}
The set of accessible states varies non-monotonically during the dynamical evolution due to non-Markovianity. Such a non-monotonic behaviour can be used as a witness for non-Markovian effects. This leads to the condition
\begin{equation}
    \frac{d V(t)}{dt}>0,
\end{equation}
to certify non-Markovianity.
The corresponding variation of accessible state volume under non-Markovian dynamics for correlated RTN, OUN and NMAD is depicted in Fig. \ref{fig: Accessible_volume}. The dynamical changes in the volume of accessible states to a system for a correlated NMAD channel are presented in Fig. \ref{fig: Accessible_volume} (a). The non-monotonic variation in volume gets enhanced with an increase in the $\mu$ value. These points to an increase in non-Markovianity with increment in channel correlation. Similar changes are observed for unital correlated RTN and are depicted in Fig. \ref{fig: Accessible_volume} (b). In the case of correlated OUN, the volume of accessible states increases with $\mu$, but no non-monotonic behaviour is observed, consistent with the findings in \cite{kumar2018non}, and is due to the CP-divisible nature of the underlying dynamics.
This is presented in Fig. \ref{fig: Accessible_volume} (c). The change in the volume of accessible states points to an enhancement in non-Markovianiaty for correlated RTN and NMAD, which are P-indivisible in nature. The increase in volume, with increment in $\mu$, is also observed for correlated OUN channel and points to the enhancement of non-Markovianity, albeit monotonic in nature, due to its inherent CP-divisiblity.

\section{\label{sec:level5i} Freezing of correlatons in correlated channels}
An interesting feature associated with correlated non-Markovian quantum channels is the possibility of freezing the dynamical evolution of certain quantum states. The study of dynamics of Bell states of the form $\ket{\phi^{\pm}}=\frac{1}{\sqrt{2}} [\ket{00} \pm \ket{11} ]$, $\ket{\psi^{\pm}}=\frac{1}{\sqrt{2}} [\ket{01} \pm \ket{10} ]$ or $X$ type states \cite{xstate}, under the action of the correlated unital non-Markovain dephasing channels (RTN, OUN) brings this out and has been observed for concurrence in Sec. \ref{sec:level4}. This freezing can also be viable for other correlation-based quantum properties like teleportation fidelity and fidelity deviation \cite{sabale}. For unital channels, the freezing phenomenon can be observed and predicted by looking at the corresponding evolved general density matrix $\rho_{u}(t)=\mathcal{E}^{corr} \rho(0)$. Therefore, the evolved density matrix can be represented as 
\begin{equation}
\rho_{u}(t)=
\begin{pmatrix}
\rho_{11} & p(t)\rho_{12} & p(t)\rho_{13}  & \tau(\mu)\rho_{14}\\
p(t)\rho_{21} & \rho_{22} & \tau(\mu)\rho_{23} & p(t)\rho_{24} \\
p(t)\rho_{31} & \tau(\mu)\rho_{32}  & \rho_{33}  & p(t)\rho_{34}\\
\tau(\mu)\rho_{41} & p(t)\rho_{42} & p(t)\rho_{43} & \rho_{44}
\end{pmatrix},
\end{equation}
where $\tau(\mu)= \mu + p(t)^{2}(1-\mu)$ and $p(t)$ is the noise function associated with channel under study as defined in Eq. (\ref{eq:probtime}). The factor $\tau(\mu)$ becomes 1 for $\mu=1$, implying freezing decay of off-diagonal terms of the density matrix. To understand this behaviour, we present the case of Bell diagonal states \cite{bdstate} having the following Bloch representation
\begin{equation}
    \rho= \frac{1}{4}(I\otimes I + \sum_{i=1}^{3}c_{i}(\sigma_{i} \otimes \sigma_{i})).
\end{equation}
The evolution of such states through correlated unital quantum channels does not change their form and the initial $\Vec{c}(0)=\{c_{1},c_{2},c_{3}\}$ is changed to the final $\Vec{c}(t)=\{c_{1} \tau(\mu),c_{2} \tau(\mu),c_{3}\}$. For the case $\mu=1$, the factor $\tau(\mu)=1$ and we observe freezing phenomenon in quantum properties as $\Vec{c}(0)=\Vec{c}(t)$ \cite{Karpat2017,frozen2}. \par

For a non-unital fully correlated NMAD channel, the generic form of the evolved state is 
\begin{widetext}
  \begin{equation}
\rho_{nu}(t)=
\begin{pmatrix}
\rho_{11} + p(t)\rho_{44} & \rho_{12} & \rho_{13}  & \sqrt{1-p(t)}\rho_{14}\\
\rho_{21} & \rho_{22} & \rho_{23} & \sqrt{1-p(t)}\rho_{24} \\
\rho_{31} & \rho_{32}  & \rho_{33}  & \sqrt{1-p(t)}\rho_{34}\\
\sqrt{1-p(t)}\rho_{41} & \sqrt{1-p(t)}\rho_{42} & \sqrt{1-p(t)}\rho_{43} & (1-p(t))\rho_{44}
\label{eq: rho_nmad}
\end{pmatrix}.
\end{equation}  
\end{widetext}
Here $p(t)$ is the noise function associated with NMAD channel as defined in Eq. (\ref{eq:pt_nmad}). The fully correlated NMAD channel is found to preserve the form of the state for Bell diagonal states with $c_{3}=-1$. The evolved state updates initial $\Vec{c}(0)=\{c_{1},c_{2},-1\}$ to $\Vec{c}(t)=\{\frac{1}{2}(c_{1} + c_{2}+(c_{1}-c_{2})(1-p(t))),\frac{1}{2}(c_{1} + c_{2}+(-c_{1}+c_{2})(1-p(t))),-1\}$. Freezing thus results in Bell diagonal states with $c_{1}=c_{2}$ and $c_{3}=-1$. A specific example of such a state is $\ket{\psi^{\pm}}$, used here. However, the states $\ket{\phi^{\pm}}$ do not show freezing behaviour in the fully correlated NMAD channel.\par
The freezing behaviour of correlated NMAD is presented in Fig. \ref{fig: Freezing} (a). The increase in $\mu$ causes the freezing of concurrence to become more predominant. The fully correlated quantum channel ($\mu=1$) exhibits complete freezing of the dynamical evolution of the above-mentioned states under the influence of unital and non-unital correlated non-Markovian channels. The freezing behaviour of correlated RTN and OUN channels is depicted in Figs. \ref{fig: Freezing} (b) and \ref{fig: Freezing} (c), respectively.

\begin{figure*}
    \subfigure[]{\includegraphics[width=0.5\linewidth]{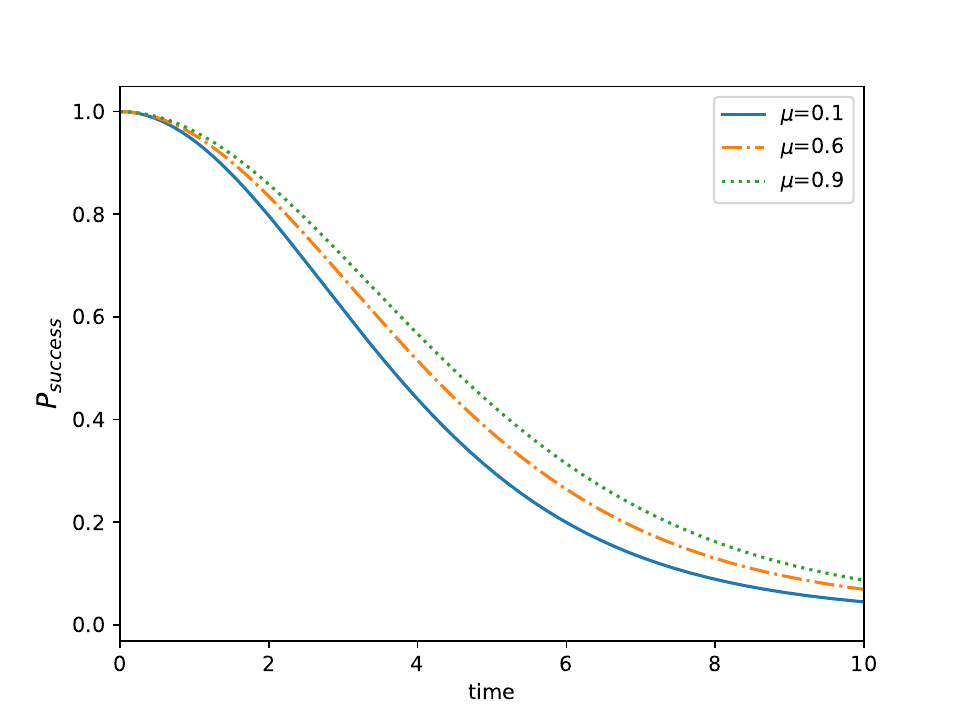}}\hfill
    \subfigure[]{\includegraphics[width=0.5\linewidth]{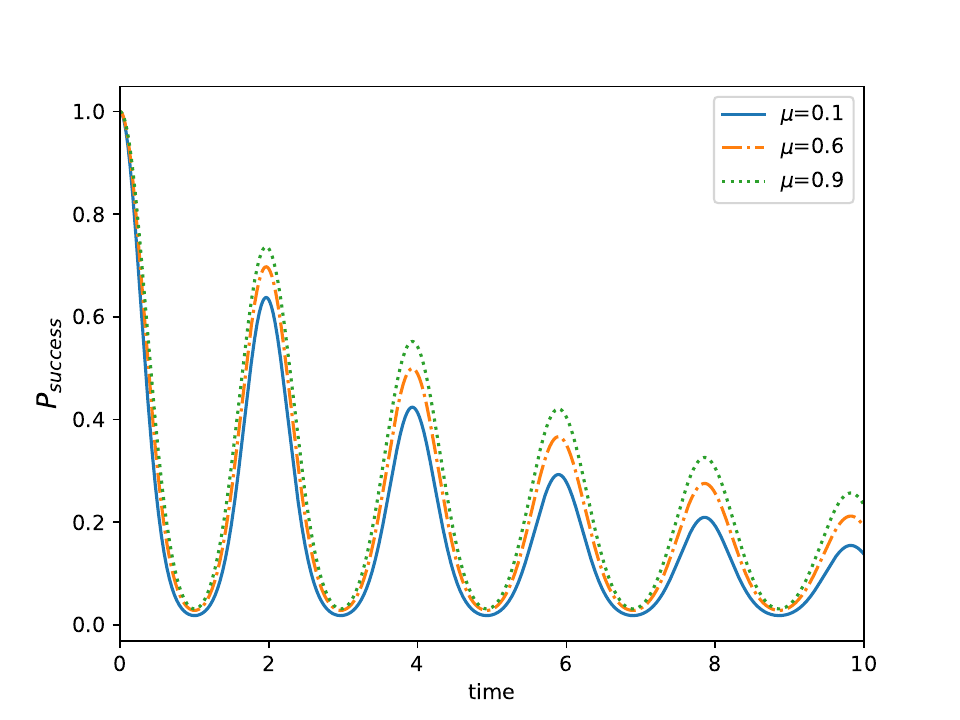}}
    \caption{The error correction success probability ($P_{success}$) of the concatenated code with (a) non-Markovian correlated OUN noise ($G=1, g=0.05$), (b)  non-Markovian correlated RTN noise ($a=0.8, \gamma=0.05 $), as a function of time ($t$) for three different values of the correlation parameter ($\mu$).}
 \label{fig: RTN_OUN_ec_success_rate}
\end{figure*}

\section{\label{sec:level5} Error correction for correlated non-Markovian dephasing channels}

It is known in the literature that correlated errors provide an advantage in quantum error correction \cite{PhysRevA.77.022323,chiribella2011quantum,clemens2004quantum}.
Here, we use quantum error correcting codes (QECCs) for a two-qubit correlated dephasing noise model (Eq. (\ref{eq:corr_map_oun_rtn})), enabling the scenario wherein both single and two-qubit dephasing errors occur. In the presented model, the channel correlation factor $\mu$ controls correlated errors. For $\mu=1$, the model exhibits fully correlated two-qubit dephasing errors. As observed above, the factor $\mu$ also has an impact on non-Markovianity. The performance of QEC for different values of $\mu$ is analysed and linked to non-Markovianity. \par

In classical and quantum error-correcting contexts, the combination of error-correcting codes results in a concatenated code \cite{forney1966concatenated,knill1996concatenated,lidar1999concatenating},
composed of an outer and inner code. The physical space of the outer code is used to construct the logical space of the inner code. The iterative use of such a concatenated code enables fault-tolerant computation by reducing error probability. This study uses concatenated quantum code to deal with correlated errors \cite{clemens2004quantum, cafaro2011quantifying, cafaro2011concatenation, dash2024concatenating}.
To approach code concatenation, we employ the three-qubit phase-flip error correcting code as an outer code to correct single-qubit dephasing errors. The encoding process here entails the transformation of a single logical qubit into a configuration of three physical qubits. The logical encoded states for this code are given as
\begin{align}
   \ket{ \Bar{0}}&=\ket{+++},\nonumber\\
    \ket{\Bar{1}}&=\ket{---}.
    \label{eq: threequbitphaseflip}
\end{align}
The stabilizer \cite{gottesman1997stabilizer} and logical operator for the three-qubit phase-flip code are $\langle XXI, XIX \rangle$, and $ZZZ$, respectively.
We use the two-qubit bit-flip code \cite{fern2008correctable} as an inner code in the concatenation, which possesses logical encoded states
\begin{align}
    \ket{\underline{0}}&=\ket{00},\nonumber\\
    \ket{\underline{1}}&=\ket{11}.
    \label{eq: twoqubitdephasing}
\end{align}
For this code, the stabilizer is $\langle ZZ \rangle$. Here, in the concatenation process, the logical states of the two-qubit bit-flip code (Eq. (\ref{eq: twoqubitdephasing})) are used to construct the physical states for the three-qubit phase-flip code (Eq. \ref{eq: threequbitphaseflip}). The encoding process follows
\begin{align}
    \ket{\pm}&=\frac{1}{\sqrt{2}}(\ket{\underline{0}} \pm \ket{\underline{1}}).
   \label{eq:plusencoding}
\end{align}
Under the two-qubit correlated dephasing error model, the concatenated QECCs can provide protection from single as well as two-qubit dephasing errors. The concatenated code resulting from the combination of outer three-qubit phase-flip code (Eq. (\ref{eq: threequbitphaseflip})) and inner two-qubit bit-flip code (Eq. (\ref{eq: twoqubitdephasing})) has logical encoded states
\begin{align}
\ket{0}_{conc}&= \frac{1}{2\sqrt{2}}\ket{000000}+\ket{000011}+\ket{001100}+\ket{001111}\nonumber\\
    &+\ket{110000}+\ket{110011}+\ket{111100},\nonumber\\
\ket{1}_{conc}&= \frac{1}{2\sqrt{2}}\ket{000000}-\ket{000011}-\ket{001100}+\ket{001111}\nonumber\\
    &-\ket{110000}+\ket{110011}-\ket{111100}.
\label{eq: logicalcodewordsconccode}
\end{align}
The form of the resultant logical state is given by
\begin{align}
\ket{\phi}_{conc}=\frac{1}{\sqrt{2}}(\ket{0}_{conc}+\ket{1}_{conc}).
\label{eq: encodedlogicalstate}
\end{align}
The resultant concatenated code can detect errors as well as correct. The error detectability condition \cite{knill2002introduction, cafaro2011quantifying} is
\begin{align}
\bra{0_{conc}}\epsilon_{i}\ket{0_{conc.}}=\bra{1_{conc}}\epsilon_{i}\ket{1_{conc}},\nonumber\\
\bra{0_{conc}}\epsilon_{i}\ket{1_{conc}}=\bra{1_{conc}}\epsilon_{i}\ket{0_{conc}}=0,
\label{eq: detectablecondition}
\end{align}
where $\epsilon_{i}$ is the set of possible errors for the encoded logical state in Eq. \ref{eq: encodedlogicalstate}. The set of correctable errors ($\epsilon_{correct}$) such that, $\epsilon_{correct}^{\dag}\epsilon_{correct}$ are detectable. The set of errors satisfying Eq. (\ref{eq: detectablecondition}) is denoted as $\epsilon_{detect}$. The details of the set of detectable and correctable errors are presented in Appendix \ref{app:qec}. One way to test the performance of the error correction is by keeping track of elements of the $\epsilon_{correct}$. Further, the error correction success probability  \cite{nielsen2010quantum,gaitan2008quantum} is defined as
\begin{align}
     P_{success}=\sum_{i} P_{correct}^{i},
\end{align}
where $P_{correct}^{i}$ is the probability of the correctable error for $i$th element from the set $\epsilon_{correct}$, refer Appendix \ref{app:qec} for details. In Fig. \ref{fig: RTN_OUN_ec_success_rate}, the error correction performance has been plotted as a function of time under OUN and RTN, in the context of non-Markovian dynamics. The respective noise dynamics can be obtained by inserting the Eqs. \ref{eq:prob}, \ref{eq:probtime}, and \ref{eq:jointprob} in the Eq. (\ref{eq:corr_map_oun_rtn}). The oscillatory nature of the RTN noise is reflected in the corresponding behaviour of the success probability. The parameter $\mu$ has been systematically varied in this analysis. It can be noticed that an increase in the parameter $\mu$ leads to a corresponding rise in the success rate. This relationship may be attributed to the retention of the correlation.
 \par

\section{\label{sec:level6} Conclusion}
The correlated channels can be a step towards newer strategies to suppress noise effects on quantum systems. The correlated non-Markovian quantum channels have been explored less and can give promising results for quantum speed limit and channel capacity related studies. \par

The present work involves constructing and analysing various types of correlated non-Markovian channels. The non-Markovianity study based on entanglement measure shows that the correlated RTN, of unital type, exhibits a decrease in memory effects for the system initially in the Bell state. The state resulting from local unitary transformation on the Bell state is found to capture an increment in non-Markovianity with an increase in channel correlation. In the case of a correlated NMAD channel of a non-unital nature, it was observed that the system starting in the Bell state exhibits an increment in non-Markovianity signatures with an increase in $\mu$. The SSS measure was studied to compute the non-Markovianity of correlated OUN channels; an increase in non-Markovianity with a higher channel-correlated factor ($\mu$) was observed. 
Enhanced non-Markovianity is further identified with state-independent measure based on a variation of the volume of physical states accessible to the system evolving through correlated quantum channels. The increment in the value of channel correlation ($\mu$) gives enhanced non-monotonic variation in the volume of physical states. The non-Markovianity of correlated OUN is not reflected in such measure due to its CP-divisible characteristics, but it shows an increase in the volume of accessible states, which decay monotonically. \par
Further study of correlated channels gives an interesting phenomenon of the freezing of correlations in time. The freezing of concurrence is observed in correlated unital channels (RTN, OUN) for the system being in the Bell state or $X$ type state. The Bloch representation of $X$ type states reveals freezing as $\Vec{c}(0)=\Vec{c}(t)$, for quantum states evolved through unital correlated quantum channels with $\mu=1$. From the perspective of freezing of correlations, the nature of dynamics in unital and non-unital correlated quantum channels differs. For correlated NMAD, freezing behaviour is observed for states with $c_{1}=c_{2}$ and $c_{3}=-1$, seen, {\it for example}, in the state $\ket{\psi^{\pm}}=\frac{1}{\sqrt{2}}\ket{01} \pm \ket{10}$. This behaviour of correlated NMAD channel can be used to distinguish the two classes of Bell states ($\ket{\phi^{\pm},\ket{\psi^{\pm}}}$), as $\ket{\phi^{\pm}}$ does not satisfy the necessary condition to exhibit freezing.\par

The exploration of correlated quantum channels and quantum error correction provides an interesting relationship between channel correlation and error correction success probability. This points to the control over noise suppression, which can be achieved by increasing the magnitude of the channel correlation factor $\mu$. The effective increase in non-Markovianity is seen to be beneficial for quantum error correction. 

\begin{acknowledgments}

VBS acknowledges the Department of Chemistry, IIT Jodhpur, and MoE for providing research facilities and financial support. NRD acknowledges financial support by the Department of Science and Technology (DST) through the INSPIRE fellowship.
\end{acknowledgments}

\appendix

\section{\label{app:all}Map to master equation and vice-versa}
The general time local master equation has the form 
\begin{equation}
    \dot{\rho}(t)= \mathcal{L}_{t}(\rho(t)),
    \label{eq:me_time_local}
\end{equation}
where $\mathcal{L}_{t}$ is a linear map acting on $\rho$. The solution of the master equation results in time-evolved density operator $\rho(t)$ 
\begin{equation}
    \rho(t)= \mathcal{E}_{t}(\rho(0))=\sum_{k} \lambda_{k}(t) A_{k}(t)\rho(0) A_{k}^{\dagger}(t),
\end{equation}
where $\lambda_{k}(t) \in \{-1,1\}$. The map $\mathcal{E}_{t}$ is completely positive for $\lambda_{k}(t)$ =1. We use tools from \cite{andersson2007finding} to relate $\mathcal{L}_{t}$ and $\mathcal{E}_{t}$ and vice versa. To define a state space of $d$ dimension, we need $N=d^{2}$ basis operators set $\{G_{m}\}$. These operators have the following properties
\begin{equation}
    G_{0}=\frac{1}{\sqrt{d}}I; \quad G_{m}= G_{m}^{\dagger} ; \quad Tr[G_{m}G_{n}]=\delta_{mn}.
\end{equation}
The Pauli operators satisfy the above-mentioned properties, making them feasible for studying single qubit channels. For two qubits, we use a basis set consisting of the following operators 
\begin{equation}
    G_{ij}=\frac{1}{2} \sigma_{i} \otimes \sigma_{j},
    \label{eq:G_basis}
\end{equation}
using the logical extension from the single-qubit case by taking the tensor product where $i,j \in \{0,4\}$. This results in sixteen basis operators for a two-qubit map. These operators are utilised to analyse two-qubit channels.

\subsection{\label{app:subsec1}Construction of $\mathcal{L}_{t}$ from given map $\mathcal{E}_{t}$}

For example, if one has a map $\mathcal{E}_{t}$ and wants to find the corresponding master equation or, in particular, $\mathcal{L}_{t}$, the procedure involves use of the above-defined basis set $G_{m}$. The use of following equation for any operator X results in an operator being expressed using a basis set $\{G_{m}\}$.
\begin{equation}
    X=\sum_{i}x_{i}G_{i}, \quad x_{i}=tr[G_{i}X].
\end{equation}
Following the same logic, the map $\mathcal{E}_{t}$ and $\rho$ are expressed as basis sums. The map takes the form 
\begin{multline}
    \mathcal{E}_{t}(\rho)=\sum_{k} tr\left[G_{k} \mathcal{E}_{t}(\sum_{l} tr[G_{l}\rho]G_{l})\right]G_{k} \\
    = \sum_{k,l} tr[G_{k}  \mathcal{E}_{t}(G_{l})] tr[G_{l} \rho] G_{k}.
    \label{eq:G_basis_map}
\end{multline}
The corresponding matrix form is given by
\begin{equation}
    \mathcal{E}_{t}(\rho)= \mathbf{(Fr)^{T}G},
\end{equation}
where 
\begin{equation}
    F_{kl}:=tr[G_{k}\mathcal{E}_{t}(G_{l})], \quad r_{l}:=tr[G_{l} \rho].
    \label{eq:F_mat}
\end{equation}
The time-evolved density operator is given by the following equation for the initial density operator $\rho(0)$
\begin{equation}
    \rho(t):= \mathcal{E}_{t}[\rho(0)].
\end{equation}
By taking the time derivative of the equation above, the resultant equation for time-dependent $\mathbf{F}$  is 
\begin{equation}
    \dot{\rho}=\mathbf{[\dot{F}r(0)]^{T}G}.
\end{equation}
Now, for $\rho(t)$ which satisfies the master equation of the form 
\begin{equation}
    \dot{\rho}=\mathcal{L}_{t}(\rho),
\end{equation}
we define a matrix $\mathbf{L}$ as
\begin{equation}
    L_{kl}:= tr[G_{k} \mathcal{L}_{t}(G_{l})],
    \label{eq:L}
\end{equation}
which for the corresponding master equation form, is given by 
\begin{equation}
    \dot{\rho}=\mathbf{[Lr(t)]^{T}G}.
\end{equation}
Then, comparing equations, we arrive at the results
\begin{equation}
\mathbf{\dot{F}r(0)}=\mathbf{Lr(t)}=\mathbf{LFr(0)}.
\end{equation}
We conclude by obtaining $\mathbf{\dot{F}=LF}$ and $\mathbf{L=\dot{F}F^{-1}}$. The matrix $L$ satisfies the relation $\norm{\mathcal{L}_{t}}= \norm{L}$. For cases where $\det \mathbf{F}=0$, one can refer to  \cite{andersson2007finding}.\par
In the case of a correlated OUN channel, the above procedure results in the following $\mathbf{L}$ corresponding to Eq. (\ref{eq:corr_map_oun_rtn}) such that
\setcounter{MaxMatrixCols}{16}
\begin{equation}
\begin{pmatrix}
0 & 0 & 0 & 0 & 0 & 0 & 0 & 0 & 0 & 0 & 0 & 0 & 0 & 0 & 0 & 0 \\
0 & \gamma & 0 & 0 & 0 & 0 & 0 & 0 & 0 & 0 & 0 & 0 & 0 & 0 & 0 & 0 \\
0 & 0 & \gamma & 0 & 0 & 0 & 0 & 0 & 0 & 0 & 0 & 0 & 0 & 0 & 0 & 0 \\
0 & 0 & 0 & 0 & 0 & 0 & 0 & 0 & 0 & 0 & 0 & 0 & 0 & 0 & 0 & 0 \\
0 & 0 & 0 & 0 & \gamma & 0 & 0 & 0 & 0 & 0 & 0 & 0 & 0 & 0 & 0 & 0 \\
0 & 0 & 0 & 0 & 0 & \gamma & 0 & 0 & 0 & 0 & 0 & 0 & 0 & 0 & 0 & 0 \\
0 & 0 & 0 & 0 & 0 & 0 & 0 & 0 & 0 & 0 & 0 & 0 & 0 & 0 & 0 & 0 \\
0 & 0 & 0 & 0 & 0 & 0 & 0 & \gamma(\mu) & 0 & 0 & 0 & 0 & 0 & 0 & 0 & 0 \\
0 & 0 & 0 & 0 & 0 & 0 & 0 & 0 & \gamma(\mu) & 0 & 0 & 0 & 0 & 0 & 0 & 0 \\
0 & 0 & 0 & 0 & 0 & 0 & 0 & 0 & 0 & \gamma & 0 & 0 & 0 & 0 & 0 & 0 \\
0 & 0 & 0 & 0 & 0 & 0 & 0 & 0 & 0 & 0 &  \gamma(\mu) & 0 & 0 & 0 & 0 & 0 \\
0 & 0 & 0 & 0 & 0 & 0 & 0 & 0 & 0 & 0 & 0 &  \gamma(\mu) & 0 & 0 & 0 & 0 \\
0 & 0 & 0 & 0 & 0 & 0 & 0 & 0 & 0 & 0 & 0 & 0 &  \gamma & 0 & 0 & 0 \\
0 & 0 & 0 & 0 & 0 & 0 & 0 & 0 & 0 & 0 & 0 & 0 & 0 &  \gamma & 0 & 0 \\
0 & 0 & 0 & 0 & 0 & 0 & 0 & 0 & 0 & 0 & 0 & 0 & 0 & 0 &  \gamma & 0 \\
0 & 0 & 0 & 0 & 0 & 0 & 0 & 0 & 0 & 0 & 0 & 0 & 0 & 0 & 0 &  0
\end{pmatrix}, 
\label{matrix:L}
\end{equation}

where,
\begin{align*}
    \gamma=-\frac{1}{2}(1-e^{-gt}),\\ 
    \gamma(\mu)=\frac{e^{-gt}(-1+e^{gt})G(-1+\mu)}{1+(-1+e^{G (\frac{-1+e^{-gt}}{g}+t)})\mu}, 
\end{align*} and the basis used for construction of $\mathbf{L}$ is given in Eq. (\ref{eq:G_basis}).

\subsection{\label{app:subsec2}Finding Kraus operators from given master equation}

For the given master equation, Eq. (\ref{eq:me_time_local}), one can compute the matrix $L(t)$ as given in Eq. (\ref{eq:L}). Further, the time-dependent $F(t)$ is computed by using the following equation

\begin{equation}
    F(t)=\mathcal{T}\exp[\int_{0}^{t} ds L(s)],
\end{equation}

where $\mathcal{T}$ is Dyson time ordering operator. In order to obtain the Kraus operators characterizing the map, from the given master equation, a useful entity needed is $\mathbf{S}$, the Choi representation of the map under study. The computation of $\mathbf{S}$ is done using the following equation

\begin{equation}
    S_{ab}=\sum_{r,s} F_{sr}tr[G_{r}\tau_{a}^{\dagger} G_{s} \tau_{b}],
\end{equation}

where $\{\tau_{a}\}$ is another basis, which could be a non-Hermitian basis. One can also resort to the computational basis $\tau_{a}= \ket{i} \bra{j}$.\par
We further consider a unitary transformation on matrix $\textbf{S}$, using the unitary operator $\textbf{W}$ so that
\begin{equation}
    S^{W}= W^{\dagger}S W.
\end{equation}
This transformation defines an associated orthonormal basis set $\{H_{a}\}$. The matrix $\textbf{W}$ and $\textbf{S}^{W}$ are given by the following equations 
\begin{equation}
    W_{ab} = tr[H_{a}\tau_{b}^{\dagger}] , \quad S_{ab}^{W} =\sum_{r,s} F_{sr}tr[G_{r} H^{\dagger}_{a} G_{s} H_{b}].
    \label{eq: sw}
\end{equation}
The matrix $\mathbf{S}$ for the master equation, Eq. (\ref{eq:two_qubit_nmad_me}), representing the correlated NMAD channel is 

\begin{widetext}
\begin{equation}
\mathbf{S}=\begin{pmatrix}
1-p(t) & 0 & 0 & 0 & 0 & \sqrt{1-p(t)} & 0 & 0 & 0 & 0 & \sqrt{1-p(t)} & 0 & 0 & 0 & 0 & \sqrt{1-p(t)} \\
0 & 0 & 0 & 0 & 0 & 0 & 0 & 0 & 0 & 0 & 0 & 0 & 0 & 0 & 0 & 0 \\
0 & 0 & 0 & 0 & 0 & 0 & 0 & 0 & 0 & 0 & 0 & 0 & 0 & 0 & 0 & 0 \\
0 & 0 & 0 & p(t) & 0 & 0 & 0 & 0 & 0 & 0 & 0 & 0 & 0 & 0 & 0 & 0 \\
0 & 0 & 0 & 0 & 0 & 0 & 0 & 0 & 0 & 0 & 0 & 0 & 0 & 0 & 0 & 0 \\
\sqrt{1-p(t)} & 0 & 0 & 0 & 0 & 1 & 0 & 0 & 0 & 0 & 1 & 0 & 0 & 0 & 0 & 1 \\
0 & 0 & 0 & 0 & 0 & 0 & 0 & 0 & 0 & 0 & 0 & 0 & 0 & 0 & 0 & 0 \\
0 & 0 & 0 & 0 & 0 & 0 & 0 & 0 & 0 & 0 & 0 & 0 & 0 & 0 & 0 & 0 \\
0 & 0 & 0 & 0 & 0 & 0 & 0 & 0 & 0 & 0 & 0 & 0 & 0 & 0 & 0 & 0 \\
0 & 0 & 0 & 0 & 0 & 0 & 0 & 0 & 0 & 0 & 0 & 0 & 0 & 0 & 0 & 0 \\
\sqrt{1-p(t)} & 0 & 0 & 0 & 0 & 1 & 0 & 0 & 0 & 0 &  1 & 0 & 0 & 0 & 0 & 1 \\
0 & 0 & 0 & 0 & 0 & 0 & 0 & 0 & 0 & 0 & 0 &  0 & 0 & 0 & 0 & 0 \\
0 & 0 & 0 & 0 & 0 & 0 & 0 & 0 & 0 & 0 & 0 & 0 &  0 & 0 & 0 & 0 \\
0 & 0 & 0 & 0 & 0 & 0 & 0 & 0 & 0 & 0 & 0 & 0 & 0 &  0 & 0 & 0 \\
0 & 0 & 0 & 0 & 0 & 0 & 0 & 0 & 0 & 0 & 0 & 0 & 0 & 0 &  0 & 0 \\
\sqrt{1-p(t)} & 0 & 0 & 0 & 0 & 1 & 0 & 0 & 0 & 0 & 1 & 0 & 0 & 0 & 0 &  1
\end{pmatrix}. 
\end{equation}
\end{widetext}
Here $p(t)$ is as in Eq. (\ref{eq:pt_nmad}). The change in basis is needed to obtain Kraus operators as one switches from the Pauli to the computational basis. The matrix $\mathbf{S^{W}}$ can be obtained using Eq. (\ref{eq: sw}), which is further used to generate Kraus operators. The corresponding map for $H_{a}=\tau_{a}$ is given by the equation
\begin{equation}
    \mathcal{E}_{t}(\rho)= \sum_{ab} S_{ab}^{w} \tau_{a} \rho \tau_{b}^{\dagger}= \sum_{i} A_{i} \rho A_{i}^{\dagger}.
\end{equation}
This map is further expressed in terms of Kraus operators. The derived Kraus operators are given in Eqs. (\ref{subeq:E0}) and (\ref{subeq:E1}).

\section{\label{app:qec} Correctable errors and error correction success probability}
The set of correlated dephasing errors for the resultant six-qubit concatenated code is
\begin{align}
\epsilon_{i}=\{I, Z\}^{\otimes 6}.
\label{eq:appcorrelatederrorsconc}
\end{align}
We want to check which elements from the above set are detectable. The error detectability condition \cite{knill2002introduction, cafaro2011quantifying} is given by
\begin{align}
\bra{0_{conc}}\epsilon_{i}\ket{0_{conc}}=\bra{1_{conc}}\epsilon_{i}\ket{1_{conc}},\nonumber\\
\bra{0_{conc}}\epsilon_{i}\ket{1_{conc}}=\bra{1_{conc}}\epsilon_{i}\ket{0_{conc}}=0,
\label{eq:appdetectablecondition}
\end{align}
where $\ket{0_{conc}}$ and $\ket{1_{conc}}$ are the logical codewords for the concatenated code in Eq. (\ref{eq: logicalcodewordsconccode}). From the set $\mathcal{E}_{i}$ (Eq. (\ref{eq:appcorrelatederrorsconc})), the set of errors that does not satisfy the error detection condition (Eq. (\ref{eq:appdetectablecondition})) for the resultant concatenated code  is 
\begin{align}
\epsilon_{undetect} & =\{ZIZIZI, ZIZIIZ, ZIIZZI, ZIIZIZ,\nonumber\\
& IZZIZI, IZZIIZ, IZIZZI, IZIZIZ\}.
\label{eq:appundetectableerrorsconc}
\end{align}
All other errors in the set $\epsilon_{i}$ are detectable.
Using Eq. (\ref{eq:appcorrelatederrorsconc}) and Eq. (\ref{eq:appdetectablecondition}), the set of correctable errors by the  concatenated code is
\begin{align}
    \epsilon_{correct}&=\bigg\{IIIIII, ZIIIII, IZIIII, IIZIII, \nonumber\\
& IIIZII, IIIIZI, IIIIIZ, ZZIIII,\nonumber\\
& IIZZII, IIIIZZ, ZZZIII, ZZIZII,\nonumber\\
& ZZIIZI, ZZIIIZ, ZIZZII, ZIIIZZ,\nonumber\\
& IZZZII, IZIIZZ,  IIZZZI, IIZZIZ,\nonumber\\
& IIZIZZ, IIIZZZ, ZZZZII, ZZIIZZ,\nonumber\\
& IIZZZZ, ZZZZZI, ZZZZIZ, ZZZIZZ,\nonumber\\
& ZZIZZZ, ZIZZZZ, IZZZZZ, ZZZZZZ \bigg\}.
    \label{eq:appcorrectabelerrorsconc}
\end{align}
Considering the first element $IIIIII$ from the set (\ref{eq:appcorrectabelerrorsconc}). Suppose we start from the rightmost side of the error element. The identity operation on the rightmost side qubit has the probability $p_{0}$. By considering the neighbouring qubit, correlation arises, and the probability of the neighboring qubit depends on the probability of the rightmost qubit. It results in the probability of the identity on the neighbour qubit of the rightmost qubit to be $p_{00}$. The same process can be followed for a six-qubit error element. Thus, the probability of the first element $IIIIII$ is calculated as
\begin{align}
    P_{correct}^{1}&= p_{00}p_{00} p_{00}p_{00} p_{00}p_{0}\nonumber\\
    &=\frac{1}{2} (1 + p(t)) \big(\frac{1}{4} (1 + p(t))^2 (1 - \mu) \nonumber\\
    &+ \frac{1}{2} (1 + p(t)) \mu \big)^5.
     \label{eq:appsuccessprob1stelements}
\end{align}
Again, considering the $14$th element $ZZIIIZ$ from the set (\ref{eq:appcorrectabelerrorsconc}), the probability is calculated following the same procedure as for the first element such that
\begin{align}
P_{correct}^{14}&= p_{33}p_{30} p_{00}p_{00} p_{03}p_{3}\nonumber\\
    &=\frac{1}{2048}\big((p(t)^2 -1)^4 (\mu-1)^2(1 + p(t) (\mu-1 ) + \mu )\nonumber\\
    & \big(1 + p(t) + \mu - p(t) \mu)^2 \big).
   \label{eq:appsuccessprob14thelements}
\end{align}
In Eqs. (\ref{eq:appsuccessprob1stelements}) and (\ref{eq:appsuccessprob14thelements}), $p_{i}$ denotes probabilities of with and without dephasing error on a single qubit given as
\begin{align}
    p_{3} &= \frac{1 - p(t)}{2},\nonumber\\
    p_{0} &= \frac{1 + p(t)}{2},
    \label{eq:appsinglequbitprob}
\end{align}
which are also referred as $q_{3}$ and $q_{0}$ respectively in Eq. (\ref{eq:prob}). The terms $p_{ij}$ are defined using the following equation 
\begin{equation}
    p_{ij}= (1- \mu) p_{i}p_{j} + \mu p_{i} \delta_{ij},
    \label{eq: jointprob}
\end{equation}
and they imply the joint probability for the correlated errors on $i$th and $j$th qubit.

In a similar fashion as Eqs. (\ref{eq:appsuccessprob1stelements}) and (\ref{eq:appsuccessprob14thelements}), we can calculate the probabilities for the rest of the elements from the set $\mathcal{E}_{correct}$. The error correction success probability is obtained by summing the probabilities of the elements in the set $\mathcal{E}_{correct}$ (Eq. (\ref{eq:appcorrelatederrorsconc})) using Eqs. (\ref{eq: jointprob}) and (\ref{eq:appsinglequbitprob}).
The total error correction success probability \cite{nielsen2010quantum, gaitan2008quantum} for the concatenated code discussed in this study is calculated as
\begin{align}
     P_{success}& =\sum_{i=1}^{32} P_{correct}^{i}\nonumber\\
   &= \frac{1}{128} \Bigg(2 + 4 p(t)^{10} (-1 + \mu)^4 + 3 \mu - \mu^{3} +\nonumber\\ 
   &p(t)^{8} \big(26 - 47 \mu + 37 \mu^{3} - 16 \mu^{4}\big)\nonumber\\ 
   & + 2 p(t)^{2} \big(10 + 11 \mu + 7 \mu^{3} + 2 \mu^{4}\big)\nonumber\\ 
   & + 2 p(t)^{6} \big(12 + \mu (-7 + 12 \mu) (-3 + \mu^{2})\big)\nonumber\\ 
   & - 4 p(t)^{4} \bigg(-13 + \mu + \mu^{2} \big(-12 + \mu (5 + 4 \mu)\big)\bigg)\Bigg),
\end{align}
where the noise functions $p(t)$ for OUN and RTN are given as
\begin{equation}
    p(t)= 
\begin{cases}
    \exp(-\gamma t)(\cos{(\omega \gamma t)}+\frac{\sin{(\omega \gamma t)}}{\omega}) & \text{for RTN} \\
    \\
    \exp[-\frac{G}{2}(t+\frac{1}{g}(\exp(-g t)-1))]              & \text{for OUN}.
\end{cases}
 \label{eq:appprobtime}
\end{equation}
\\
\\

\bibliography{final}
\end{document}